\documentclass{icrc2009}

\usepackage{graphicx}   % for including figures
\usepackage{caption}    % for captions
\usepackage[font=footnotesize]{subfig} % subfig.sty for a double column floating figure using two subfigures
\usepackage{fixltx2e}
\usepackage{url}

\newcommand{\shorttitle}[1]%
{\markboth{Proceedings of the 31\MakeLowercase{$^{st}$} ICRC, {\L}\'{o}d\'{z} 2009}{#1} }
\newcommand{\etal}{\MakeLowercase{\textit{et al. }}} % "et al."

\begin{document}
\title{Sensitivity of JEM-EUSO to GRB Neutrinos}

\author{\IEEEauthorblockN{Katsuaki Asano\IEEEauthorrefmark{1},
			  Kenji Shinozaki\IEEEauthorrefmark{2} and
                          Masahiro Teshima\IEEEauthorrefmark{3}\\
for the JEM-EUSO collaboration}
                            \\
\IEEEauthorblockA{\IEEEauthorrefmark{1}Graduate School of Science,
Tokyo Institute of Technology, 2-12-1 Ookayama
Meguro-ku, Tokyo 152-8550, Japan}
\IEEEauthorblockA{\IEEEauthorrefmark{2}Comutational Astrophysics Laboratory,
RIKEN Advanced Science Institute, 2-1 Hirosawa, Wako 351-0198, Japan}
\IEEEauthorblockA{\IEEEauthorrefmark{3}Max-Planck-Institute for Physics,
Foehringer Ring 6, D-80805 Munich, German}}

% please write the preseter's name and short title (3-4 words maximum)
%    which will appear at the header of the even pages.
\shorttitle{Asano \etal JEM-EUSO and GRB neutrinos}
\maketitle

\begin{abstract}
JEM-EUSO is a mission to study ultra-high-energy cosmic rays (UHECRs) by measuring the fluorescence light from giant air showers at the altitude of the International Space Station. In the tilted mode, JEM-EUSO will become very sensitive to the \v{C}erenkov light from the earth skimming tau neutrinos at the energy range of $10^{16-18}$eV. In this paper we will discuss high-energy tau neutrinos from nearby gamma-ray bursts (GRBs). From simulations of cascade in GRB photon fields including various
hadronic/leptonic processes, we estimate the neutrino flux from GRBs. Our results show that
both muons and pions are dominant sources of neutrinos
at the energy range of $10^{16-18}$ eV.
We discuss the possibility of detecting the \v{C}erenkov light of upward going
showers from Earth skimming tau neutrinos coming from some closest GRBs.
\end{abstract}

 
\section{Introduction}

JEM-EUSO (Extreme Universe Space Observatory) \cite{euso} is
an observatory that will be installed on the Japanese Experiment Module (JEM)
on the International Space Station (ISS) at an altitude of approximately 400km.
The launch is planned for 2013 by H2B rocket and conveyed by the HTV
(H-II transfer Vehicle) to ISS.
It observes extensive air showers (EAS) induced by ultra-high-energy cosmic rays (UHECRs)
with the energy higher than about $10^{19}$ eV.
The super-wide Field-of-View (FOV $\pm 30^\circ$) yields a circle with a 250 km radius
of the observational aperture of the ground area.
The targets of JEM-EUSO are not only UHECRs but also upward neutrino events.
The atmospheric volume in the FOV of JEM-EUSO is about $10^{12}$ tons.
However, the effective area is increased by inclining the telescope from nadir (tilted mode),
so that the target volume for upward neutrino events exceeds $10^{13}$ tons.
On the other hand, the threshold energy to detect EAS increases too.
The first half of the mission lifetime is devoted to observing the lower energy region
in the nadir mode, and the second half of the mission to observing the high energy region
in tilted mode.

The physical conditions of gamma-ray bursts (GRBs)
inferred for internal shocks \cite{mes06} indicate that
protons may be Fermi-accelerated to energies of $\sim 10^{20}$ eV,
making GRBs potential sources of
the observed UHECRs \cite{wax95}.
To test this GRB  UHECR scenario,
it is indispensable to search for UHE proton-induced signatures of
secondary neutral radiation
that can be observed in coincidence with GRBs.
Although the proton-induced gamma-ray emissions, such as
synchrotron emission from secondary electrons/positrons
injected via photomeson interactions,
have been discussed frequently \cite{der06,asa07},
it may not be easy from photon spectra
to distinguish the signature of protons from
inverse Compton emissions due to primary accelerated electrons.
Therefore, the detection of high-energy neutrinos from GRBs
\cite{wax97} is essential to verify the proton acceleration in GRBs.
Many authors have estimated the cosmological neutrinos background
due to GRBs \cite{mur08}.
The correlation between neutrino signals and GRBs may verify a GRB UHECR scenario.
JEM-EUSO provides us the unique chance to detect very high-energy neutrinos
above $10^{16}$ eV, but whose background flux is probably too low to correlate
each shower event with a GRB.
Alternatively, we discuss the possibility of neutrino detection
from a nearby bright GRB with JEM-EUSO.

\section{Model}

We employ the numerical code in Asano et al. (2009) \cite{asa09}
(see also \cite{asa07}) to
simulate hadronic cascade processes in GRBs with Monte Carlo techniques.
While Asano et al. (2009) focused on the photon emissions due to accelerated protons,
the numerical code automatically outputs resultant neutrino spectra.
The physical processes taken into account are 1) photon emission
processes of synchrotron and Klein-Nishina regime Compton scattering
for electrons/positrons, protons, pions, muons, and kaons,
2) synchrotron self-absorption for electrons/positrons,
3) $\gamma \gamma$ pair production, 4) photomeson production from protons,
5) photopair production ($p \gamma \to p e^+ e^-$) from protons,
6) decays of pions, muons, and kaons.
For photomeson productions, the method is the same as
Asano \& Nagataki (2006) \cite{asa06};
we adopt experimental results for the
cross sections of
$p (n) \gamma \to n \pi^+ (\pi^0)$,
$p \pi^0 (\pi^-)$, $n \pi^+ \pi^0 (\pi^-)$,
and $p (n) \pi^+ \pi^-$, while theoretically obtained values \cite{dre92,lee01}
are adopted for kaon productions via
$p \gamma \to \Lambda K^+$, $\Sigma^0 K^+$, and $\Sigma^+ K^0$,
or $n \gamma \to \Lambda K^0$, $\Sigma^- K^+$, and $\Sigma^0 K^0$.
More details on the treatment of meson production and their decay products
can be found in Asano (2005) \cite{asa05} and Asano \& Nagataki (2006) \cite{asa06}.

Both electrons and protons are assumed to be injected with power-law energy distributions
as $\propto E^{-p} \exp{(-E/E_{\rm max})}$.
In order to make a GRB UHECR scenario viable,
the proton spectrum index is required to be $p_{\rm p}=2$, which is harder
than the typical index of electrons $p_{\rm e}$ obtained from GRB photon spectra.
Although the injection spectra for the two species
are expected to be the same at low energies where their gyroradii overlap,
$p_{\rm e} > p_{\rm p}$ may be effectively realized
if the proton spectrum covering 7-8 decades in energy deviates from a pure power-law
and becomes concave.
Such possibilities are shortly mentioned in Asano et al. (2009)
\cite{asa09}.
Hence, we adopt $p_{\rm p}=2$ and $p_{\rm e}=2.5$ in our simulations as optimistic cases.
The maximum proton energy is determined by
equating the acceleration timescale and cooling/dynamical timescale,
taking into account synchrotron, IC, and photomeson cooling.

We consider relativistically expanding shells, where
all photons and particles are distributed isotropically
in the shell frame and treated in the one-zone approximation.
We use conventional notation $\epsilon_{\rm e}$, $\epsilon_{\rm p}$ and $\epsilon_{\rm B}$,
for the energy fractions of accelerated electrons, protons and magnetic fields
to the shock-dissipated internal energy.
The minimum Lorentz factor of electrons
is chosen so that the corresponding synchrotron photon energy
is always 300 keV.
Thus, the full set of our model parameters consists of
the bulk Lorentz factor $\Gamma$, variability timescale $\Delta t=R/c \Gamma^2$
(or emission radius $R$), energy of accelerated electrons in the shell $E_{\rm sh}$,
the energy fractions $\epsilon_B/\epsilon_{\rm e}$ and $\epsilon_{\rm p}/\epsilon_{\rm e}$.
In the plasma rest frame,
we follow all physical processes mentioned above in a volume $4 \pi R^3/\Gamma$
during a timescale $R/c\Gamma$ with sufficiently short time steps.

 \begin{figure}[!t]
  \centering
  \includegraphics[width=3.0in]{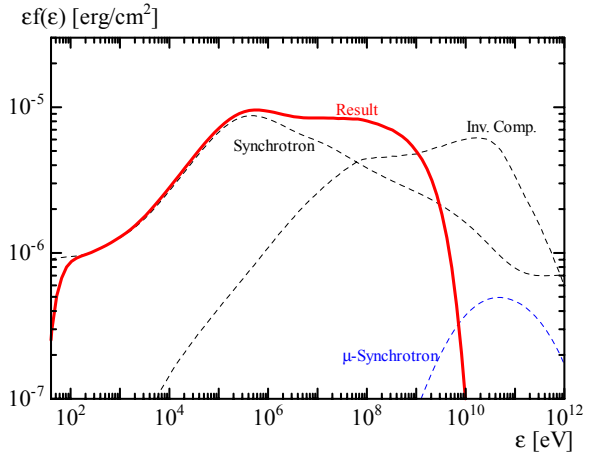}
  \caption{Photon spectrum (bold solid curve)
of a GRB pulse at 1 Gpc of the luminosity distance ($z=0.2$) with $\Gamma=500$,
$\Delta t=50$ ms, $E_{\rm sh}=10^{52}$ erg, $\epsilon_B/\epsilon_{\rm e}=0.1$,
and $\epsilon_{\rm p}/\epsilon_{\rm e}=1.0$.
Fine dashed curves denote separately electron/positron synchrotron,
inverse Compton, and muon synchrotron components
without the absorption effects.}
  \label{phs}
 \end{figure}

Fig. \ref{phs} shows the photon spectrum obtained by our simulation
assuming the equipartition between protons and electrons,
$\epsilon_{\rm p}/\epsilon_{\rm e}=1.0$. The other parameters
are denoted in the figure caption. The proton cascade
processes enhance the photon energy by $\sim 7$\%.
We can clearly see the effects of $\gamma \gamma$ absorption and
synchrotron self-absorption around GeV and 100 eV energies,
respectively.
The deviation from a simple power-law spectrum
near the synchrotron self-absorption energy is
due to the heating via self-absorption,
which modifies the effective electron distribution
(the synchrotron boiler effect \cite{ghi88}).
The contribution of inverse Compton emission is suppressed
because of the Klein-Nishina effect in spite of the
lower value of $\epsilon_B/\epsilon_{\rm e}=0.1$.

As shown in Fig. \ref{phs}, a significant amount of muons
is produced via photomeson production.
Since our code outputs escaped photons and neutrinos simultaneously,
the resultant neutrino spectra are consistent
with obtained photon spectra, taking into account
cooling of pions, muons and kaons.

\section{Neutrino Detection Sensitivity}

\begin{figure}[!t]
  \centering
  \includegraphics[width=3.0in]{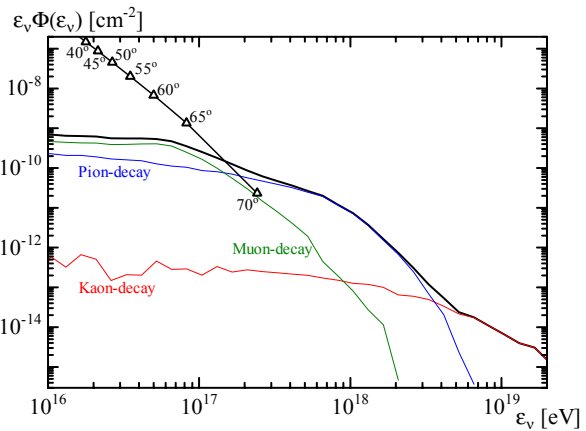}
  \caption{$\tau$-neutrino spectra (bold solid curve) for the same parameter set
as Fig. \ref{phs}. 100 identical pulses are assumed to be emitted from a GRB
at 1 Gpc (the total energy of gamma-rays $\sim 10^{54}$ erg).
Fine solid curves denote neutrinos from decay of muons,
pions, and kaons as labeled.
The triangles show the threshold energy and the 99\% C.L. flux sensitivity with JEM-EUSO
as a function of the nadir angle as labeled.
}
  \label{nut}
 \end{figure}

Assuming the neutrino cross section $\sigma_{\nu {\rm N}}
=2.0 \times 10^{-33} (\varepsilon_\nu/10^{16} \mbox{eV})^{0.363} \mbox{cm}^2$,
$\tau$-deacy length $500 (E_\tau/10^{16} \mbox{eV})$ m,
reflective index of air $n=1.0003$ (the \v{C}erenkov light cone of $1.4^\circ$),
and density of the mantle $4.5 \mbox{g}/\mbox{cm}^3$,
we estimate the sensitivity and threshold energy of $\tau$-neutrinos
as a function of the nadir angle
for JEM-EUSO at an altitude of 400km (see Fig. \ref{nut}).

Since almost all neutrinos in the nadir direction are absorbed within the earth,
it may be very difficult to detect EAS due to $\tau$-decay in the nadir mode
(the nadir angle of neutrinos $<30^\circ$).
As the nadir angle of neutrinos in the FOV increases,
the threshold energy of neutrinos gets higher,
since the mean distance to EAS and atmospheric absorption both increase.
However, the larger nadir angle and resultant higher energy threshold lead to
a larger target volume due to the longer lifetime of $\tau$ in the Earth crust,
which give us a effectively thicker target volume.
The larger nadir angle also lead to a longer distance between tau showers and the detector which produces a larger light pool of Cerenkov and effectively gives a wider target area on the Earth.  These two effects give us a huge effective target volume for the detection of showers induced by tau neutrinos.
As a result, the sensitivity increases with the nadir angle
of neutrinos.
In this estimate we have not performed detailed Monte Carlo simulations of
development of EAS etc., which is in preparation now,
so that the sensitivity curve in the figure is preliminary.

In Fig. \ref{nut}, we plot $\tau$-neutrino spectra,
assuming that neutrino oscillation makes
the fraction of tau-neutrinos 1/3.
As Asano \& Nagataki (2006) \cite{asa06} pointed out,
neutrinos from kaons dominate in the highest energy range above $10^{19}$ eV,
because kaons, heavier than pions, do not lose so much energy
before they decay into neutrinos.
In the energy range of $10^{17}$-$10^{18.5}$ eV,
the dominant component is neutrinos from $\pi^\pm$-decay,
while neutrinos from $\mu$-decay is the primary component below $10^{17}$ eV.
The maximum energies of neutrinos are determined by the balance between
the cooling time and the decay time of seed particles.
The obtained spectrum shows a complex shape reflecting the difference
in the cooling of seed particles.

The feasible tilt angle of JEM-EUSO may be below $35^\circ$,
so that a nadir angle should be $<65^\circ$.
Even for this nearby (1 Gpc) and bright
(total gamma-ray energy $\sim 10^{54}$ erg) GRB,
we need $>65^\circ$ for the nadir angle of neutrinos,
though more sophisticated estimates via EAS simulations may
significantly improve the sensitivity.

\section{Proton-Dominated GRBs}

While the equipartition between protons and electrons is assumed
in the former section,
we should consider the possibility
that GRBs contain a significantly larger amount of energy in protons
compared to that radiated by the accelerated electrons.
From the local UHECR emissivity at proton energy $\sim 10^{19}$ eV,
the necessary isotropic-equivalent energy per burst
in accelerated protons integrated over $\sim 10^9 - 10^{20}$ eV
is $\sim 2 \times 10^{54} - 3 \times 10^{55}$ erg \cite{gue07}.
On the other hand, isotropic-equivalent
gamma-ray energy is
typically $\sim 10^{53}$ erg and up to $\sim 10^{54}$ erg.
Thus, in order for GRBs to be viable sources of UHECRs,
the latest observations point to a highly proton-dominated energy budget,
$\epsilon_{\rm p}/\epsilon_{\rm e} \sim 10$-100,
which is approximately independent of the actual beaming factor.
In the internal shock model,
shocks convert a fraction of the bulk kinetic energy of outflows
into Fermi-accelerated relativistic electrons.
Initially, however, most of the kinetic energy as well as
the internal energy generated via shock dissipation
are likely carried by protons,
The simple 1$^{\rm st}$ order Fermi acceleration theory may predict higher
acceleration efficiency of protons compared to electron acceleration,
because the Larmor radius of shock-dissipated electrons is smaller than that of protons.

 \begin{figure}[!t]
  \centering
  \includegraphics[width=3.0in]{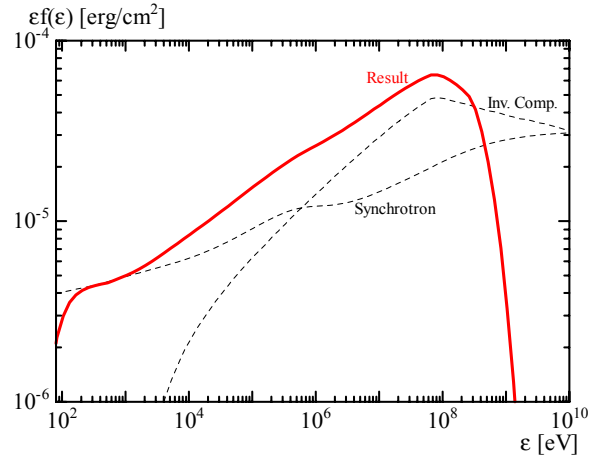}
  \caption{Same as Fig. \ref{phs} but for
$\epsilon_{\rm p}/\epsilon_{\rm e}=30$.
}
  \label{phs-pd}
 \end{figure}

 \begin{figure}[!t]
  \centering
  \includegraphics[width=3.0in]{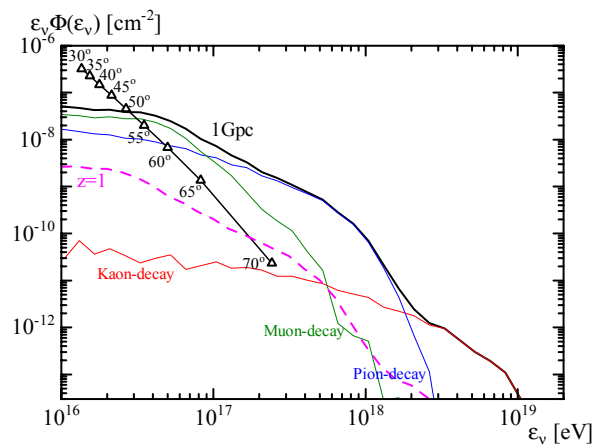}
  \caption{Same as Fig. \ref{nut} but for
$\epsilon_{\rm p}/\epsilon_{\rm e}=30$.
The bold solid line is for a GRB at 1 Gpc,
while the dashed line is for $z=1$.
}
  \label{nut-pd}
 \end{figure}

As Asano et al. (2009) \cite{asa09} supposed,
we calculate photon and neutrino spectra for proton-dominated GRBs
as shown in Figs. \ref{phs-pd} and \ref{nut-pd}.
In our parameter choice $\epsilon_{\rm p}/\epsilon_{\rm e}=30$,
the proton-induced secondary emission totally overwhelms any primary 
electron component, resulting in a hard spectrum peaking at 100 MeV,
which is determined by $\gamma \gamma$ optical depth.
The dominant components are inverse Compton and synchrotron emission
from secondary electrons/positrons.
The proton contribution enhances resultant gamma-ray energy by a factor of 5.
Although the obtained photon spectrum differs from the typical GRB spectra,
Kaneko et al. (2008) \cite{kan08} reported a similar GRB spectrum
with a high peak energy $> 170$ MeV,
as well as a few other GRBs with significant high-energy excess.
Some studies have also indicated potential observational biases against detections
of high peak energies \cite{llo99}.
If we adopt a larger $\Gamma$, the proton cascade efficiency is diminished,
which can lead to a typical shape of GRB spectra.
However, lower efficiency of the proton cascade weakens neutrino emission, too.

The $\tau$-neutrino spectrum for such an optimistic case is plotted
in Fig. \ref{nut-pd}.
The larger amount of protons and higher efficiency of pion production
amplify the neutrino flux. As shown in Fig. \ref{nut-pd},
if such a nearby and bright GRB with the nadir angle $>50^\circ$
occurs inside the FOV of JEM-EUSO, EAS due to $\tau$-neutrinos will be detected.
For the nadir angle of $\sim 70^\circ$, more distant GRBs at $z \sim 1$ can be
detected.

\section{Discussion}

If a nearby ($< 1$ Gpc) GRB fortunately occurs,
a considerable neutrino flux has been predicted by our model
for the nadir angle $\sim 65^\circ$.
Post-{\it SWIFT} estimates of the local rate of long GRBs
range from $0.2-1\ {\rm Gpc^{-3} yr^{-1}}$ \cite{gue07},
if the GRB rate is proportional to the star formation rate.
Therefore, we may expect a few GRBs within 1 Gpc per year.
For a tilt angle of $35^\circ$,
neutrinos from the nadir angle $65^\circ$ yield
a very tiny solid angle in the FOV.
If we can take a larger tilt angle, though it may be not realistic,
the solid angle for the nadir angle $65^\circ$-$70^\circ$ can be $\sim 0.1$ rad.
So the detection possibility of nearby GRBs may be a few percent per year in this case,
though we should take into account the dead time of observation.
If proton-dominated GRBs at 1 Gpc,
which is compatible with
the GRB-UHECR scenario, are actual cases,
neutrinos with the nadir angle $>50^\circ$
can be detected. The solid angle of such neutrinos
is $\sim 0.1$ rad in the FOV for the tilted mode of $35^\circ$,
which implies a few percent of the GRB detection possibility per year.
Although the above estimate of the GRB rate in the FOV may sound pessimistic,
those finite possibilities are not negligible.
The luminous neutrino flux in the proton-dominated model encourages us
to detect neutrinos from GRBs even at $z \sim 1$,
which is the typical redshift of GRBs ($\sim 100$ events per year).
In this optimistic scenario,
the larger volume within $z \sim 1$
enhances the detection possibility, while we need a larger tilt angle.

A more realistic sensitivity curve for upward neutrino events
is now in preparation by the JEM-EUSO collaboration.
The unique capability of JEM-EUSO provides us with an exclusive
chance to detect very high-energy neutrinos from GRBs.

\end{document}